\documentstyle [prl,aps]{revtex}

\begin{document}

\title{Bianchi Type I space and the
Stability of Inflationary FRW solution}

\author{ W.F. Kao\thanks{email:wfgore@cc.nctu.edu.tw}}
\address{Institute of Physics, Chiao Tung
University, Hsin Chu, Taiwan}
\maketitle

\begin{abstract}
Stability analysis of the Bianchi type I universe in pure gravity theory is studied in details.
We first derive the non-redundant field equation of the system by introducing the generalized Bianchi type I metric.
This non-redundant equation reduces to the Friedmann equation in the isotropic limit.
It is shown further that any unstable mode of the isotropic perturbation with respect to a de Sitter background is also unstable with respect to anisotropic perturbations.
Implications to the choice of physical theories are discussed in details in this paper.
\end{abstract} \vskip .2in


\section{Introduction}
Inflationary theory provides an appealing resolution for the the flatness, monopole, and horizon problems of our present universe described by the standard big bang cosmology \cite{inflation}. 
It is known that our universe is homogeneous and isotropic to a very high degree of precision \cite{data,cobe}.
Such a universe can be described by the well known Friedmann-Robertson-Walker (FRW) metric \cite{book}. 

It is also known that gravitational physics should be different from the standard Einstein models near the Planck scale \cite{string,scale}. 
For example, quantum gravity or string corrections could lead to interesting cosmological consequences \cite{string}.
Moreover, some investigations have addressed the possibility of deriving inflation from higher order gravitational corrections \cite{jb1,jb2,Kanti99,dm95}.

A general analysis of the stability condition for a variety of pure higher derivative gravity theories is very useful in many respects. 
In fact, it was shown that a stability condition should hold for any potential candidate of inflationary universe in the flat Friedmann-Robertson-Walker (FRW) space \cite{dm95}. 

In addition, the derivation of the Einstein equations in the presence of higher derivative couplings is known to be very complicate.
The presence of a scalar field in induced gravity models and dilaton-gravity model makes the derivation even more tedious.
In order to simplify the complications in the derivation of the field equations, an easier way has been described \cite{kp91,kpz99}.
We will try to generalize the work in Ref. \cite{kpz99} in order to obtain a general and model-independent formula for the non-redundant field equations in the Biachi type I (BI) anisotropic space. 
This equation can be applied to provide an alternative and simplified method to obtain the stability conditions in pure gravity theories.
In fact, this general and model-independent formula for the non-redundant field equations is very useful in many area of interests.
In particular, it will be applied to study a large class of pure gravity models with inflationary BI$/$FRW solutions in this paper.
Any Bianchi type I solution that leads itself to an asymptotic FRW metric at time infinity will be referred to as the BI$/$FRW solution in this paper for convenience.

Note that there is no particular reason why our universe is initially isotropic to such a high degree of precision.
Even anisotropy can be smoothed by the proposed inflationary process, it is also interesting to study the stability of the FRW space during the post-inflationary epoch.
Nonetheless, it is natural to expect that our universe starts out as an anisoptropic universe.
The universe is then expected to evolve from certain anisotropic universe, e.g. a Bianchi type I universe, to an isotropic universe, such as the flat FRW space.
Indeed, it was shown that there exists such kind of BI$/$FRW solution for a NS-NS model with a metric field, a dilaton and an axion field \cite{CHM01}.
This inflationary solution is also shown to be stable against small field perturbations \cite{ck01}.
Note also that stability analysis has been studied in various fields of interest \cite{kim,abel}.

A large class of models with the BI$/$FRW solutions will be shown to be unstable against arbitrary anisotropic perturbations in this paper.
We will first derive a stability equation which turns out to be identical to the stability equation  for the existence of the inflationary de Sitter solution discussed in Ref. \cite{dm95,kpz99}.
Note that an inflationary de Sitter solution in pure gravity models is expected to have one stable mode and one unstable mode for the system to undergo inflation with the help of the stable mode.
Consequently, the inflationary era will come to an end once the unstable mode takes over after a brief period of inflationary expansion.
The method developed in Ref. \cite{dm95,kpz99} was shown to be a helpful way in choosing physically acceptable model for our universe.
Our result indicates, however, that the unstable mode will also tamper the stability of the isotropic space.
To be more specific, if the model has an unstable mode for the de Sitter background perturbation with respect to isotropic perturbation, this unstable mode will also be unstable with respect to any anisotropic perturbations.

\section{Non-redundant field equation and Bianchi identity}

Note that the generalized Bianchi type I (GBI) metric can be read
off directly from the following equation:
\begin{equation}
ds^2 \equiv g^{\rm GBI}_{\mu\nu} dx^\mu dx^\nu =  -b^2 (t) dt^2 + {a_1^2}(t) dx^2
+a_2^2(t) dy^2 + a_3^2 (t) dz^2,
\label{eqn:gbi} \label{GBI}
\end{equation}
with $b(t)$ the lapse function restored on purpose.
Note also that the Bianchi type I (BI) metric can be obtained from GBI metric by setting
the lapse function  $b(t)$ equal to one, i.e. $b=1$, in equation
(\ref{eqn:gbi}).

Note that one can list all non-vanishing components of the curvature tensor as
\begin{eqnarray}
R^{ti}_{\,\,\,\,tj}&=&{1\over 2} [H_i \dot{B}+2B(\dot{H}_i+H^2_i)]\delta^i_j
\label{Rti}\\
R^{ij}_{\,\,\,\,kl}&=& H_iH_j B  C^{ij}_{\,\,\,\,kl}   \label{Rkl}.
\end{eqnarray}
Here $C^{ij}_{\,\,\,\,kl} \equiv \epsilon^{ijm} \epsilon_{mkl}$ with
$\epsilon^{ijk}$ the three space Levi-Civita tensor \cite{book}. 
Here $\dot{}$ denotes differentiation with respect to $t$ and
$H_i= \dot{a}_i/a_i$ is the directional Hubble constant.
We have also written $B \equiv 1 / b^2$ for later convenience.
Note also that the indices $i,j$ in both sides of above equations are open indices.
Moreover, $R^{ij}_{\,\,\,\,kl}=0$ if $i=j$ due to the symmetric properties of the curvature tensor.

Given a pure gravity model one can cast the action of the system as 
$S=\int d^4x \sqrt{g} {\cal L}=N \int dt (a_1 a_2 a_3 / \sqrt{B} ) L(H_i, \dot{H}_i, B, \dot B) $ in the GBI spaces. 
Here $N$ is a time independent integration constant. 
We will denote the volume factor as $V\equiv a_1 a_2 a_3 b $.
If we take $VL$ as an effective Lagrangian, one can show that the variation with respect to $b$ gives
\begin{eqnarray} \label{key0}
-{1 \over 2} L + {1 \over 2} H_iL_i  +\dot{H}_i L^i -{1 \over 2} (\dot H _i +3 H H_i +H_i d/dt)L^i=0 
\end{eqnarray}
after setting $B=1$.
Here the last term comes from the variation with respect to $\delta (H_i \dot B)$ with the help of the identity
\begin{equation} \label{thekey}
{\delta L \over \delta H_i \dot B}  \delta (H_i \dot B) =
{\delta L \over \delta 2 B \dot H_i} \delta (H_i \dot B) \to
{\delta L \over \delta 2 \dot H_i} H_i  \delta \dot B
\end{equation}
once we set $B=1$.
One also needs an integration-by-part with respect to $\delta B$ in order to obtain the result indicated in the last term of Eq. (\ref{key0}).
Note that the first equality in the Eq. (\ref{thekey}) comes from the fact that the factor $H_i \dot B$ always shows up with $2 B \dot H_i$ as indicated by the curvature tensor $R^{ti}_{\;\;\;tj}$ in Eq. (\ref{Rti}).
Therefore, one ends up with the $b$-equation as
\begin{eqnarray} \label{beq} \label{key}
L-H_iL_i = (\dot H _i -3 H H_i -H_i d/dt)L^i 
\end{eqnarray}
after setting $B=1$.
Here $H \equiv \sum_i H_i /3$, $L_i \equiv \delta L /\delta H_i$  and $L^i \equiv \delta L /\delta \dot H_i$ for convenience.
In addition, one can also derive,
\begin{equation}
L+ (3H + d /dt)^2 L^i = (3H + d /dt) L_i, \label{aieq}
\end{equation}
as the variational equation of $a_i$.
The derivation of this equation is tedious but straightforward.
In addition, $VL$ is normally referred to as the effective Lagrangian.
We will also call $L$ as effective Lagrangian unless confusion occurs.
Note that the equations (\ref{aieq}) are in fact the space-like $ij$ component of the Einstein equation
\begin{equation}
G_{\mu\nu} = t_{\mu\nu}          \label{einstein}
\end{equation}
with $t_{\mu\nu}$ denoting the generalized energy momentum tensor associated
with the system. 
It is known that one of these equations is in fact a redundant equation. 
Indeed, one can define $H_{\mu\nu} \equiv G_{\mu\nu} - t_{\mu\nu}$ and write the field equation as $H_{\mu\nu}=0$.

Hence one has
\begin{equation}
D_\mu H^{\mu\nu} =0 \label{DH}
\end{equation}
from the energy conservation ($D_\mu t^{\mu \nu}=0$) and the
Bianchi identity ($D_\mu G^{\mu \nu}=0$).
Now we have three independent scale factors $a_i$ and four equations.
Therefore, one of the four equations has to be redundant.
Indeed, the extended Bianchi identity (\ref{DH}) can be shown to give
\begin{equation}
(\partial_t + 3 H) H_{tt} + \sum_i H_i H_{ii}=0, \label{h3}
\end{equation}
as soon as the BI metric is substituted into equation (\ref{DH}).
Therefore, the equation (\ref{h3}) indicates that: ``$H_{tt}=0$
implies $\sum_i H_i H_{ii}=0$".
Hence, two of the $H_{ii}$ equations vanish will imply that the third one also vanishes.

On the other hand, $H_1=H_2=H_3=0$ implies instead
$(\partial_t + 3H ) H_{tt}=0.$
This implies that $V H_{tt} = {\rm constant}$ with $V \equiv a_1a_2a_3$.
Hence the $H_{tt}$ equation is the non-redundant equation while we are
free to ignore one of the three $H_{ii}$ equation.
Hence any conclusion derived without the $H_{tt}$ equation is known to be incomplete.

\section{perturbation and stability}
One can then apply the perturbation, $H_i=H_{i0}+ \delta H_i$, to the field equation with $H_{i0}$ the background solution to the system.
This perturbation will enable one to understand whether the background solution is stable or not.
In particular, one would like to learn whether a BI$/$FRW type evolutionary solution is stable or not.
It is known that our universe could start out anisotropic even evidences indicate that our universe is isotropic to a very high degree of precision in the post inflationary era.
Therefore, one expects that any physical model should admit a stable BI$/$FRW solution.


Our result indicates that FRW inflationary solutions with a stable mode and an unstable mode is a negative result to our search for a physically acceptable model.
Note that  FRW inflationary solutions with a stable mode and an unstable mode will provide a natural way for the inflationary universe to exist the inflationary phase.
Our result indicates, however, that such models will also be unstable against the anisotropic perturbations.
Therefore, such solution will be harmful for the system to settle from BI space to FRW space once the graceful exist process is done.

First of all, one can show that the first order perturbation equation from the non-redundant field equation (\ref{key}), with $H_i \to H+\delta H_i$, gives
\begin{eqnarray}
L_{02} H \delta  \ddot{H}_i
+ \left [ (3H^2 -\dot H )L_{02} + 3H \dot H L_{12} + 3H \ddot H L_{03}  \right ] \delta \dot{H}_i \nonumber && \\
+(6H L_{01} +3H^2 L_{11} -HL_{20}+\ddot H L_{02} + 3 H \ddot H L_{12} + 3 H \dot H L_{21} ) \delta H_i =0  &&
\label{stable1}
\end{eqnarray}
where all field variables are understood to be evaluated at the background FRW space where $H_i = H = \dot a /a$ for all directions, with $a$ the FRW scale factor.
We will write $H$ as the Hubble parameter for the FRW space for convenience from now on.
$L_{ab} \equiv \delta^{a+b} L / \delta H_{i_i} \delta H_{i_2} \cdots H_{i_a} \delta \dot H _{j_1} \delta \dot H_{j_2}
\cdots \delta \dot H _{j_b} |_{H_i \to H}$.
In deriving above equations, we have used the following identities
\begin{eqnarray}
\sum_i H_i {\delta L \over \delta H_i}  &\to& H {\delta L \over \delta H}, \\
{\delta L \over \delta H_i}  &\to&  {\delta L \over 3\delta H}, \\
f( \sum_i H_iH_i) &\to & f(3H^2)
\end{eqnarray}
when we take the limit $H_i \to H$ carrying the system from the BI space to the flat FRW space limit.
Here $f (\sum_i H_iH_i)$ denotes any functions of the variable $\sum_i H_iH_i$.
One can show that the Eq. (\ref{stable1}) reduces to
\begin{eqnarray}
L_{02} \delta  \ddot{H}_i
+ 3H_0 L_{02} \delta \dot{H}_i 
+(6 L_{01} +3H_0 L_{11} -L_{20}) \delta H_i =0 \label{stable0}
\end{eqnarray}
once we set $H=H_0=$ constant which denotes the de Sitter space.
This equation is identical to the stability equation for the existence of an inflationary de Sitter solution discussed in Ref. \cite{dm95,kpz99}.
One notes that there are two other independent $H_{ij}$-equation remain to be checked.
These equations can be shown to be redundant after the limiting case $H_0=$ constant is implemented. 
Indeed, the anisotropic perturbation on the $H_{ij}$-equations are expected to reproduce the redundant isotropic perturbation equation shown in Ref. \cite{kpz99}.

Note that an inflationary de Sitter solution is expected to have one stable mode and one unstable mode for the system to undergo inflation with the help of the stable mode.
Indeed, one can show that the Eq. (\ref{stable0}) can be solved to give
\begin{equation}
\delta H_i = c_i \exp [B_+ t]
+ d_i \exp [B_- t]
\end{equation}
with $B_\pm = -3H_0/2 \pm \sqrt{\Delta_0}/2 L_{02}$ and arbitrary constants $c_i, d_i$ to be determined by the initial perturbations.
Here $\Delta_0 \equiv 9H_0^2 L_{02}^2 -4 L_{02} (6L_{01}+3H_0 L_{11}- L_{20})$
is the discriminant of the characteristic equation for the Eq. (\ref{stable0})
\begin{equation}
L_{02} x^2
+ 3H_0 L_{02} x  
+(6 L_{01} +3H_0 L_{11} -L_{20}) =0 . \label{stable00}
\end{equation}
There will be an unstable mode if $B_+ >0$.
Therefore, the inflationary era will come to an end once the unstable mode takes over.
It was shown earlier to be a helpful way to select physically acceptable model for our universe.
Our result shown here indicates, however, that the unstable mode will also tamper the stability of the isotropic space.
Indeed, if the model has an unstable mode for the de Sitter perturbation, this unstable mode will also be unstable against the anisotropic perturbation.

For example, one can show that the model \cite{dm95}
\begin{equation}
{\cal L} = -R -\alpha R^{\mu \nu}_{\:\:\: \beta \gamma} \, R^{\beta \gamma}_{\:\:\: \sigma \rho} \,
R^{\sigma \rho}_{\:\:\: \mu \nu}
\end{equation}
admits an inflationary solution when $\alpha <0$.
Note that this model is the minimal consistent effective low-energy two-loop renormalizable Lagrangian for pure gravity theory \cite{2loop}.
Indeed, one can show that 
\begin{equation}
L= \sum_i \left \{ 
2\dot H_i + 4 H_i^2 
- 4 \alpha \left [ 2 (\dot H_i + H_i^2)^3 - H_i^6 \right ]  \right \}
- 4 \alpha (\sum_i H_i^3)^2
\to 6 (\dot H + 2 H^2) -24  \alpha \left [ (\dot H + H^2)^3 +H^6 \right ]
\end{equation}
when we set $H_i \to H$.
Hence one can show that the generalized Friedmann equation (\ref{key}) gives
\begin{equation}
H_0^4 = -1/4 \alpha .
\end{equation}
In addition, the stability equation (\ref{stable1}) for $\delta H_i$ can be shown to be
\begin{equation}
12 \alpha H_0^2 \delta \ddot H_i
+36 \alpha H_0^3 \delta \dot H_i 
-(1+ 12 \alpha H_0^4 ) \delta H_i
=0
\end{equation}
This equation can be solved to give
\begin{equation}
\delta H_i = c_i \exp [(\sqrt{35/3}-3)H_0t/2 ]
+ d_i \exp [-(\sqrt{35/3}+3)H_0t/2 ]
\end{equation}
with arbitrary constants $c_i, d_i$ to be determined by the initial perturbations.
This indicates that this model admits one stable mode and one unstable mode following the stability equation (\ref{stable0}) for the inflationary de Sitter solution.
It is shown to be a positive sign for an inflationary model that is capable of resolving the graceful exist problem in a natural manner.
Our result indicates however that this model also admits an unstable mode against anisotropic perturbation.
Hence this model will have problem remaining isotropic for a long period of time.
Therefore, pure gravity model of this sort will not solve the graceful exist problem.
One will need, for example, the help of certain scalar field to end the inflation in a consistent way.

One expects any unstable mode for a model to be of the form
$\delta H_i \sim \exp [lH_0t]$, to the lowest order in $H_0t$, in a de Sitter background with $l$ some constant characterizing the stability property of the model.
In such models, the inflationary phase will only remain stable for a period of the order
$\Delta t \sim 1/lH_0$.
The inflationary phase will start to collapse after this period of time.
This means that the de Sitter background fails to be a good approximation when $t \gg \Delta t$.
Hence the anisotropy will also grow according to
$\delta H_i \to \delta H_i^0 \exp [lH_0 \Delta t]$ with $\delta H_i^0$ denoting the initial perturbation.

The measure of anisotropy can be estimated by computing the anisotropy parameter
$A \equiv \sqrt{\sum_i \delta H_i^2/3H^2} \sim
 \sqrt{\sum_i (\delta H^0_i)^2/3H_0^2} \exp [lH_0t]$.
Hence $A \sim \sqrt{\sum_i (\delta H^0_i)^2/3H_0^2} \exp[60l]$ for a typical inflationary model which requires a $60$ $e$-fold expansion.
This gives us a hope that small anisotropy observed today can be generated by the initial inflationary instability for models with appropriate factor $l$.

On the other hand, it is known that the Einstein-Hilbert model
\begin{equation}
{\cal L} = -R -2 \Lambda
\end{equation}
admits only stable modes, which requires $\delta H_i=0$, which is bad for the natural graceful exist.
This model is, however, stable against anisotropic perturbations which tends to keep the universe isotropic as long as the model is in charge.

\section{conclusion}
In short, the result of this paper shows that graceful exist and stability of any de Sitter model can not work along in a naive way.
The physics behind the inflationary de Sitter models appears to be much more complicate than one may expect.
In another words, the phase transition during and after the inflationary phase deserves more attention and requires extraordinary care in order to resolve the problem lying ahead.

Note that the non-redundant field equation for field theories with many different sorts of fields coupled to the system can also be derived similar to the pure gravity models \cite{kpz99}.
Similar arguments also apply to these theories. 
Therefore, theories with one unstable mode under anisotropic metric perturbations with respect to the de Sitter background will not be able to hold the de Sitter space stable for a long period of time even there exist stable modes.
Note that, in general, one needs to consider two different fourth derivative curvature terms when higher derivative theories are considered in four dimensions.
 
\vspace{0.5in}

\section{Acknowledgments}
This work was supported in part by the National Science Council of Taiwan under the contract number NSC89-2112-M009-043.

\end{document}